# Inspecting the use of SLMs for the control of photonic quantum states

# Análisis del uso de SLMs para el control de estados cuánticos fotónicos

S. Bordakevich[1,2], D. Pabón[1,2], L. Rebón[3,4], S. Ledesma[1,2,*]

*1. Universidad de Buenos Aires, Facultad de Ciencias Exactas y Naturales, Departamento de Física. Buenos Aires, Argentina.*

*2. Consejo Nacional de Investigaciones Científicas y Técnicas (CONICET), Argentina.*

*3. Departamento de Ciencias Básicas, Facultad de Ingeniería, Universidad Nacional de La Plata, La Plata, Argentina.*

*4. Instituto de Física de La Plata, UNLP - CONICET, Argentina.*

(*) E-mail: ledesma@df.uba.ar

**ABSTRACT:**

Spatial light modulators (SLMs) are widely used to coherently control quantum states of light. When carrying out these experiments, some assumptions are made. For instance, it is supposed that the position-momentum correlations between twin photon pairs are not affected by the use of a liquid crystal display (LCD) as a SLM. Furthermore, it is assumed that the characterization of such devices performed with an intense laser source, is still valid in the single photon regime. In this work, we show that such assumptions are acceptable, within the experimental uncertainties, for a liquid crystal on silicon (LCoS) display. This is especially important when considering the use of this kind of displays for the coherent control of quantum states based on twin photon sources.

**RESUMEN:**

Los moduladores espaciales de luz (SLM) son ampliamente utilizados para el control coherente de estados cuánticos de la luz. Cuando se realizan estos experimentos, suelen hacer algunas suposiciones. Por ejemplo, se asume que el uso de una pantalla de cristal líquido (LCD) actuando como SLM no afecta las correlaciones posición-momento de pares de fotones gemelos. Además, se supone que la caracterización del SLM realizada con una fuente láser intensa sigue siendo válida en el régimen de un solo fotón. En este trabajo, se muestra que ambas suposiciones son aceptables dentro de las incertidumbres experimentales, para una pantalla de cristal líquido sobre silicio (LCoS). Esto es de particular importancia al considerar el uso de este tipo de pantallas para el control coherente de estados cuánticos basados en fuentes de fotones gemelos.

**Palabras clave:** MODULADORES ESPACIALES DE LUZ, FUENTES DE FOTONES GEMELOS, CORRELACIONES ESPACIALES

# 1. Introduction

Liquid crystal displays (LCDs) have been widely used as spatial light modulators (SLMs), since they can be exploited to modulate different properties of the electromagnetic field. Additionally, they are relatively inexpensive and widely available, which makes them an attractive option for several experimental applications. Particularly, they have been used in different quantum optics experiments, and applications to quantum communication and quantum information processing [1–6].

In those experiments, correlations between properties of the photons, for example, polarization correlation, angular position and OAM correlation, or position and transverse momentum correlation, play an important role. But, in particular, the impact on position-momentum correlations [7] due to the pixel-by-pixel change of the gray level of an LCD has not been, to the best of our knowledge, systematically analyzed in the literature.

As shown in Refs. [8,9], a direct detection in the near and far fields of the photons allows to distinguish between quantum and classical correlations, in the continuous variables of momentum and position, for systems of pairs of photons. For instance, when working with spontaneous parametric down-conversion (SPDC) photon sources, position-momentum correlations come from the fact that twin photons are generated in the same place, and the transverse linear momentum of the system is conserved [10]. For photons labeled as 1 and 2, these correlations can be quantified by $P(x_2|x_1)$, the conditional probability density function of detecting photon 2 in position $x_2$, given that photon 1 was detected in position $x_1$; and $P(p_2|p_1)$, the conditional probability density function of detecting photon 2 with momentum $p_2$, given that photon 1 was detected with momentum $p_1$. Under that description, the uncertainties of those distributions should be small, so the product of the variances of the probability density functions $\Delta^2 x_2|_{x_1} \Delta^2 p_2|_{p_1}$ is usually used as a measure of position-momentum correlation degree [9], where:

$$\Delta^2 x_2|_{x_1} = \int x_2^2 P(x_2|x_1) dx_2 - \left( \int x_2 P(x_2|x_1) dx_2 \right)^2 \qquad (1)$$

and

$$\Delta^2 p_2|_{p_1} = \int p_2^2 P(p_2|p_1) dp_2 - \left( \int p_2 P(p_2|p_1) dp_2 \right)^2. \qquad (2)$$

Since the same applies for photon 1 given the detection of photon 2, labels can be switched, and $\Delta^2 x_1|_{x_2} \Delta^2 p_1|_{p_2}$ is an equally valid measure of position-momentum correlation degree. For instance, this measure represents a realization of the Einstein-Podolski-Rosen (EPR) paradox [11]. If the twin photon pair exhibit EPR-type correlations $\Delta^2 x_2|_{x_1} \Delta^2 p_2|_{p_1} \leq \hbar^2/4$ [12], which are stronger than any classical correlation, the joint state must be a non-separable state; i.e., photons are entangled. This measure has been used to study the influence of pump coherence in the entanglement of down-converted photons [13,14], in quantum communications protocols [15], to characterize quantum channels [16], and to certify entanglement for telecommunications applications [17]. It has also been applied with the usage of 2D detectors [18–21] and sources other than SPDC [22]. In a wider sense, position-momentum correlations have been used for the generation of spatial qudits [1–4], quantum light shaping [5] and quantum information processing [6].

Then, in order to coherently manipulate the quantum state of photons, it is key that the correlations of interest do not degrade when interacting with the optical elements and LCDs that are part of the experimental setup. In addition, it is also important that the characterization of such displays is performed properly.

In quantum optics experiments, single photons or twin photons are commonly used, so the light intensity regime with which they are conducted is extremely low. This experimental issue sometimes leads to the assumption that the experimental setup works in the single photon regime in the same way as when working in the intense light regime. For example, it is often assumed that the Mueller matrix of the optical elements is the same in both regimes, thus characterizing them with intense light and then using these results when working with single photons. This is understandable since the Mueller matrix of every optical element is necessary, they have to be obtained for every experimental situation (e.g., wavelength of the source, orientation of the optical element), and several measurements have to be done to obtain them. Moreover, to perform this characterization using single photons is much more time-consuming due to the needed integration times, and also alignment becomes a much more difficult task under such low light conditions.

Particularly, when characterizing LCDs, one of the most important dependences is that with the gray level of the display: a voltage difference drives the orientation of the liquid crystal molecules [23], leading to a change in the polarimetric properties of the screen. Thus, every gray level has to be described by a different Mueller matrix. This means longer characterization times and reinforces the benefits of using intense light for that task.

In this work, we address both questions for a reflective liquid crystal on silicon (LCoS) display: its ability to preserve spatial momentum-position correlations of twin photon pairs, while modulating their properties of interest; and the possible discrepancies in the Mueller matrices obtained from a characterization with such source from those obtained with an intense laser source.

## 2. Position-momentum correlations

In this section, we describe the experimental setup used to measure both the position and the momentum conditional probability density functions of down-converted photons. This was achieved by registering when a photon was detected, given that its twin was detected in a different, but known, position and having different, but known, momentum. Then, a LCoS display was introduced, and its effect on those correlations was studied. Those experiments are explained in sections 2.a and 2.b, respectively.

In both cases, a diaphragm shapes a 405 nm laser diode beam that is then collimated, resulting in a beam with a full width at half maximum of (1.13 ± 0.02) mm and a power of ~15 mW. It pumps a 10 mm thick $LiIO_3$ nonlinear crystal, generating down-converted photons. They are each directed towards their respective photon detection system, which consist of an adjustable width slit with micrometric resolution, 10 nm interferometric bandpass filters around 810 nm, and lenses that focus the photons into single mode optical fibers. Each detection system is mounted on micrometric positioners. The collected photons are then detected by Perkin Elmer SPCM-AQRH-13-FC single photon counting modules (SPCM). Those detections are postselected by a FPGA board programmed to find simultaneous detections in 4 ns temporal slots. By doing this, every coincidence detection is known to be triggered by a pair of twin photons generated in the crystal. However, some coincidences are spurious, as statistically some unrelated photons can arrive to both detectors at the same time: this amount of accidental coincidences was estimated [24] and substracted for every measurement. They were obtained by averaging several consecutive measurements, lasting 1 second each; consequently, the uncertainties are the standard error of that mean, which is independent of the count statistics. However, these uncertainties were compared to the square root of the total number of coincidences, obtaining that they exhibit a Poisson-like behavior. Schematics of the setups used without and with the LCoS display can be seen in Fig. 1 and Fig. 4, respectively.

## 2.a. Free space propagation

As can be seen in Fig. 1, the detection systems were located 106 cm apart from the crystal, and maximum coincidence detection rate was found when the detection systems were ~3.3° from the direction of the pump. The micrometric positioners allowed to move one collector in a 2.5 cm wide range while the other one was kept fixed at the position with the maximum coincidence rate, thus obtaining the probability of measuring a photon given that its twin was measured in a different, but known, position.

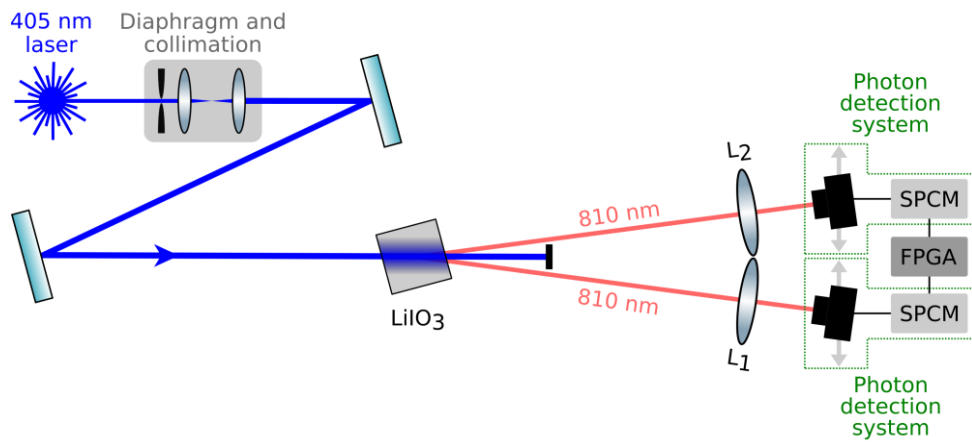

Fig. 1. Schematic of the experimental setup used to measure the position and momentum correlations with free space propagation. A 405 nm laser beam pumps a $LiIO_3$ nonlinear crystal. Photon detection systems collect the down-converted photons, to be detected by single photon counting modules (SPCM), and postselected by a FPGA board. Lenses $L_1$ and $L_2$ allow switching between position and momentum detection.

The position distribution was obtained by imaging the crystal onto the plane of the detector with lenses $L_1$ and $L_2$. The coincidence rate measured for each position in that plane was scaled according to the magnification of each image. Specifically:

- $L_1$ had a focal length of 200 mm and was located at (87.9 ± 0.2) cm from the detection systems, thus resulting in a magnification of (4.861 ± 0.014). The slits were set to ∼100 μm width, and the measurements were made by positioning the detection systems every (100 ± 3) μm.
- $L_2$ had a focal length of 150 mm and was located at (79.2 ± 0.2) cm from the detection systems, thus resulting in a magnification of (2.962 ± 0.012). The slits were set to ∼150 μm width, and the measurements were made by positioning the detection systems every (150 ± 3) μm.

In this condition, the maximum coincidence rate was around 40/s and measurements were averaged for 30 s. The resulting normalized probability density functions can be seen in Fig. 2, labeled as "Free space".

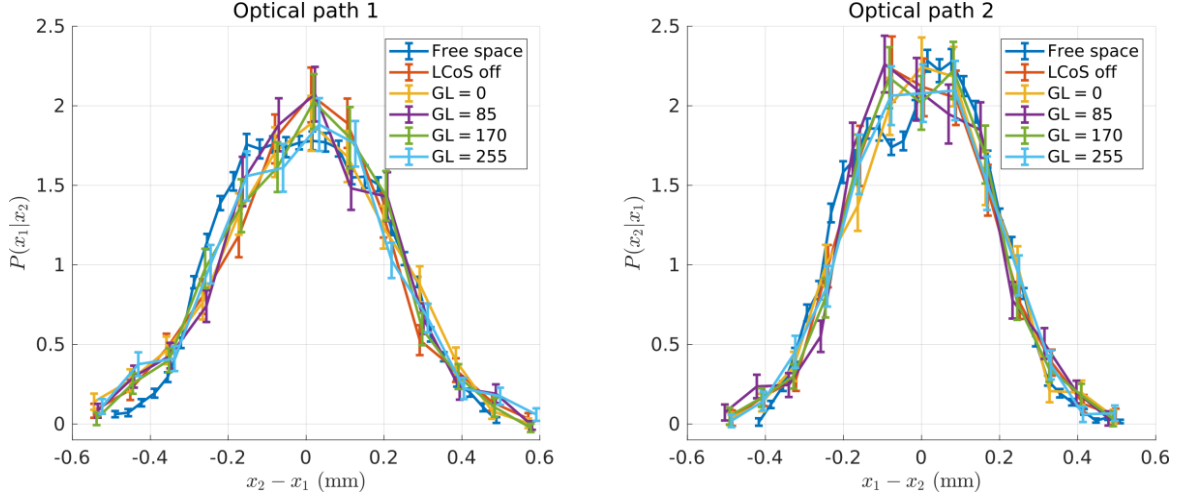

Fig. 2. Position normalized probability density functions, for each optical path. Setup in Fig. 1 was used for the free space propagation curve, and setup in Fig. 4 for the curves involving the LCoS display. The corresponding variances are in Table 1.

On the other hand, the momentum distribution was obtained by locating lenses $L_1$ and $L_2$ so its Fourier plane is in the plane of the detector. The coincidence rate measured for every position in that plane was scaled according to the focal length of the lens. Specifically:

- $L_1$ had a focal length of 200 mm and was located at (20.0 ± 0.2) cm from the detection systems, thus resulting in a scaling factor of $2\pi/\lambda f \sim 38.8$ mm$^{-2}$. The slits were set to ∼50 μm width, and the measurements were made by positioning the detection systems every (25 ± 3) μm.
- $L_2$ had a focal length of 150 mm and was located at (15.0 ± 0.2) cm from the detection systems, thus resulting in a scaling factor of $2\pi/\lambda f \sim 51.7$ mm$^{-2}$. The slits were set to ∼50 μm width, and the measurements were made by positioning the detection systems every (25 ± 3) μm.

In this condition, the maximum coincidence rate was around 60/s and measurements were averaged for 30 s. The resulting normalized probability density functions can be seen in Fig. 3, labeled as "Free space".

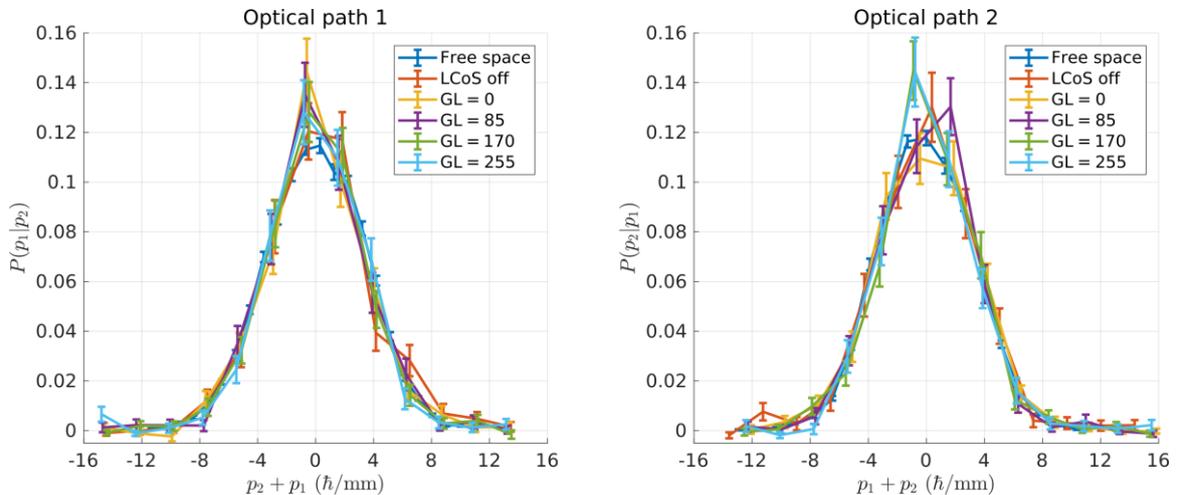

Fig. 3. Linear momentum normalized probability density functions, for each optical path. Setup in Fig. 1 was used for the free space propagation curve, and setup in Fig. 4 for the curves involving the LCoS display. The corresponding variances are in Table 1.

The variance of every set of measurements was calculated, and thus the variance product was calculated for each optical path of the setup. The result was that, for path 1, $\Delta^2 x_1|_{x_2}\Delta^2 p_1|_{p_2} = (0.41 \pm 0.07)\hbar^2$, and for path 2, $\Delta^2 x_2|_{x_1}\Delta^2 p_2|_{p_1} = (0.29 \pm 0.05)\hbar^2$. Those differences are usual and can be explained as differences in the shape of the down-converted beams, which are influenced by the pump transverse profile [7], and by beam waist and crystal length [25].

## 2.b. Influence of an LCoS display

To study the influence of a liquid crystal display in those variances, a HOLOEYE LC-R 2500 reflective liquid crystal on silicon (LCoS) display was added to the setup. This can be seen in Fig. 4, which also shows the use of two non-polarizing beamsplitters to redirect the photons to the detection systems. This means the loss of 3 out of 4 photons, thus difficulting the measurements, but it was necessary due to the aperture of ~3.3° of the down-converted beams. Additionally, the screen also causes losses.

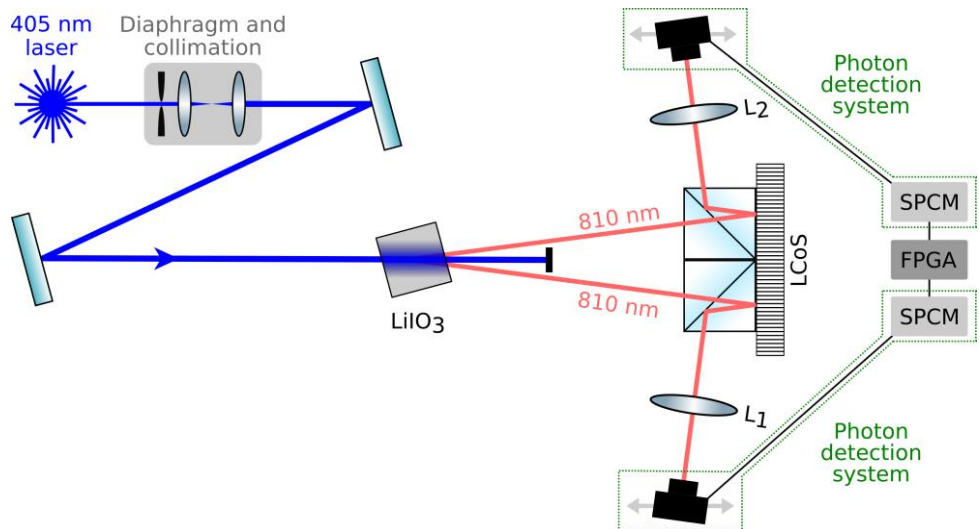

Fig. 4. Schematic of the experimental setup used to measure the position and momentum correlations with a LCoS display. A 405 nm laser beam pumps a LiIO$_3$ nonlinear crystal. Photon detection systems collect the down-converted photons, to be detected by single photon counting modules (SPCM), and postselected by a FPGA board. Lenses L$_1$ and L$_2$ allow switching between position and momentum detection.

The measurement of the position and momentum distributions was made in a similar way as the case in which the photons are allowed to propagate freely. In every case, the lenses were located between the beamsplitters and the detection systems.

For the position distribution:

- L$_1$ had a focal length of 100 mm and was located at (13.2 ± 0.2) cm from the detection systems, thus resulting in a magnification of (0.323 ± 0.016). The slits were set to ~75 μm width, and the measurements were made by positioning the detection systems every (30 ± 3) μm.
- L$_2$ had a focal length of 125 mm and was located at (18.6 ± 0.2) cm from the detection systems, thus resulting in a magnification of (0.488 ± 0.012). The slits were set to ~100 μm width, and the measurements were made by positioning the detection systems every (40 ± 3) μm.

In this condition, the maximum coincidence rate was ~1.3/s and measurements were averaged for 120 s.

For the momentum distribution:

- L$_1$ and L$_2$ had a focal length of 100 mm and were located at (10.0 ± 0.2) cm from the detection systems, thus resulting in a scaling factor of $2\pi/\lambda f \sim 77.6$ mm$^{-2}$. The slits were set to ~75 μm width, and the measurements were made by positioning the detection systems every (30 ± 3) μm.

In this condition, the maximum coincidence rate was ~0.80/s and measurements were averaged for 120 s.

The display was used in five different configurations: disconnected from its energy source (thus acting mainly as a mirror), and in four of its 256 gray levels: 0, 85, 170 and 255. The resulting normalized probability density functions can be seen in Fig. 2 and Fig. 3, labeled as "LCoS off" and "GL = 0, 85, 170, 255", respectively.

As can be seen in those figures, the widths of the probability density functions seem not to change significantly according to the configuration of the display. This is confirmed when comparing the variance products, presented in Table 1. The uncertainties are high, reaching almost 20%, which is expected given the error bars in Fig. 2 and Fig. 3. Those errors are due to the measurements being not too long in time. The uncertainties reach almost 20%, which is expected given the error bars in Fig. 2 and Fig. 3. A way to lower the uncertainties is to increase the measuring times, because in our experiment the coincidence count rates are low. This can be attributed to several factors: the relatively low pump power of 15 mW, the fact that total counts decrease by a factor of ~9 due to the display and the beamsplitters, and the ratio between coincidences and total counts of <0.1% despite fine-tuning of the alignment.

TABLE 1. Product of variances of position and momentum probability density functions in Fig. 2 and Fig. 3, for every configuration and optical path.

| Configuration | $\Delta^2 x_1|_{x_2} \Delta^2 p_1|_{p_2}$ ($\hbar^2$) | $\Delta^2 x_2|_{x_1} \Delta^2 p_2|_{p_1}$ ($\hbar^2$) |
|---|---|---|
| Free space propagation | 0.41 ± 0.07 | 0.29 ± 0.05 |
| LCoS off | 0.49 ± 0.09 | 0.30 ± 0.05 |
| Gray level 0 | 0.46 ± 0.09 | 0.32 ± 0.05 |
| Gray level 85 | 0.42 ± 0.08 | 0.31 ± 0.06 |
| Gray level 170 | 0.39 ± 0.07 | 0.29 ± 0.05 |
| Gray level 255 | 0.43 ± 0.07 | 0.27 ± 0.05 |

## 3. Mueller matrices

In this section, we address the question of if the Mueller matrices of the LCoS display are the same for intense light and for single photons. To do so, we used two experimental setups, that allowed to calculate the Mueller matrices of the HOLOEYE LC-R 2500 display for those two cases. This was achieved by using a Mueller polarimeter, seen in Fig. 5 and Fig. 7. It consisted of a polarization state generator (including a Glan–Thompson polarizer $P_G$ and a quarter-wave plate $QW_G$) before the display, and a polarization state analyzer (including a quarter-wave plate $QW_A$ and a Glan–Thompson polarizer $P_A$) after it. It was located at one optical path, so it analyzed half of the display with light impinging at ~3.3° from its normal. The photon detections on the other path were used to herald the arrival of their twins. In that condition, 49 intensity measurements were performed: both polarizers were fixed, while $QW_G$ and $QW_A$ were rotated to 7 different angles between 0 and $2\pi$, multiples of ~51.43°. From those measurements, the 16 elements of the Mueller matrix were obtained, following the procedure in Ref. [26]. This was performed for 26 gray levels, starting with GL=0 and in steps of 10 gray levels. The value of those elements as a function of gray level can be seen in Fig. 6 (a), labeled as "SPDC".

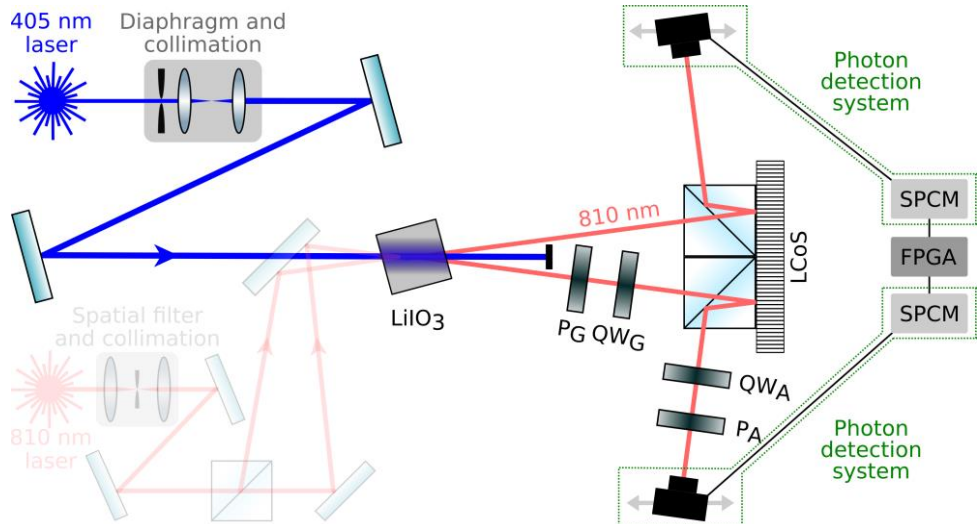

Fig. 5. Schematic of the experimental setup used to measure the Mueller matrices of a LCoS display. A 405 nm laser beam pumps a LiIO$_3$ nonlinear crystal. A Mueller polarimeter (polarizers $P_G$ and $P_A$, and quarter-wave plates $QW_G$ and $QW_A$) analyzes the down-converted photons. Photon detection system collect them to be detected by single photon counting modules (SPCM), and postselected by a FPGA board.

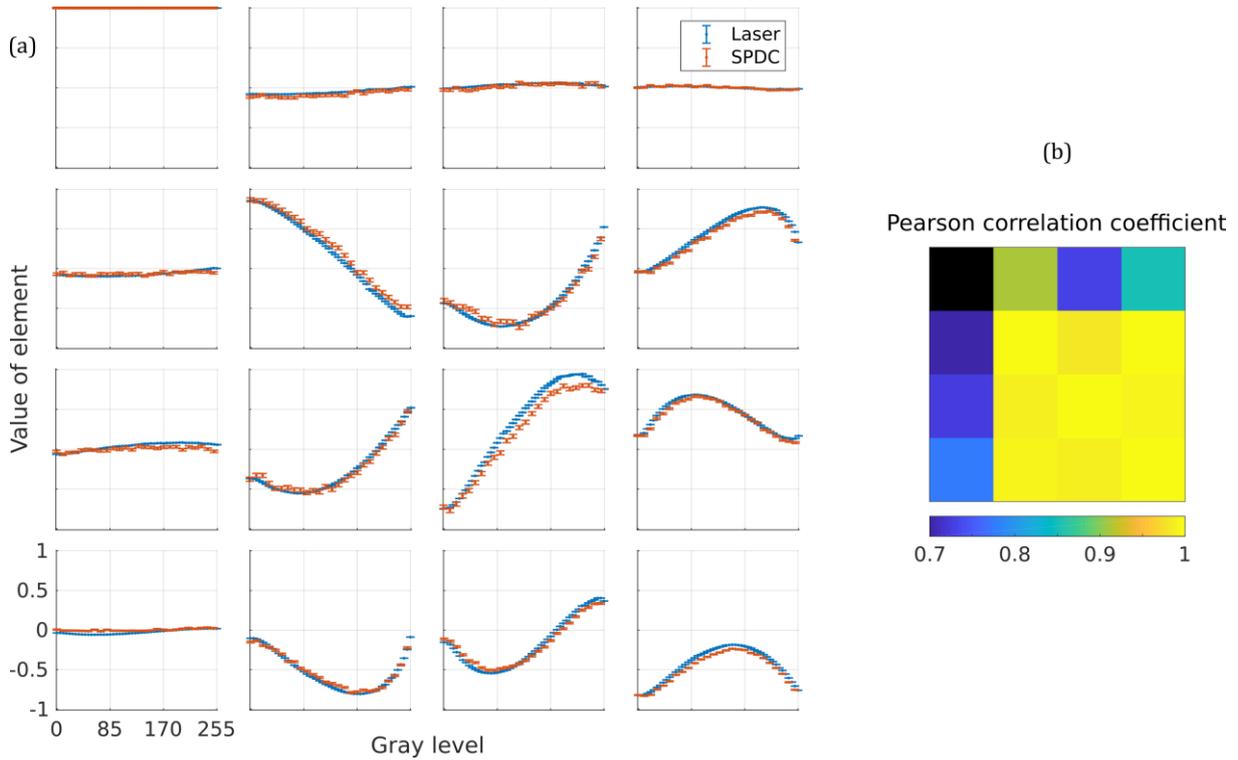

Fig. 6. (a) Value of elements of the Mueller matrix of the LCoS display as a function of its gray level, with its corresponding errors. In orange, measured with SPDC-originated single photons, in blue measured with an intense laser source. Axes and scales are the same for every element, and are shown in the bottom left corner. The top left element is 1 for every gray level due to normalization. The used setups are in Fig. 5 and Fig. 7, respectively. (b) Pearson correlation coefficient between the curves of the matrix elements. Top left is coloured black due to normalization in (a).

To perform a similar measurement but using intense light, an 810 nm laser diode was incorporated to the setup, as can be seen in Fig. 7. By means of a beamsplitter and mirrors, the beam was directed towards the crystal in a way that mimics the path followed by the down-converted photons. The polarimeter and the display were kept in place, while the photon detection system was replaced by a Newport 1918-R power meter. Without the need to herald the arrival of photons to the detector, the other path was ignored. The elements of the Mueller matrices were calculated in the same way, but for 52 gray levels in steps of 5. The obtained results are in Fig. 6, labeled as "Laser".

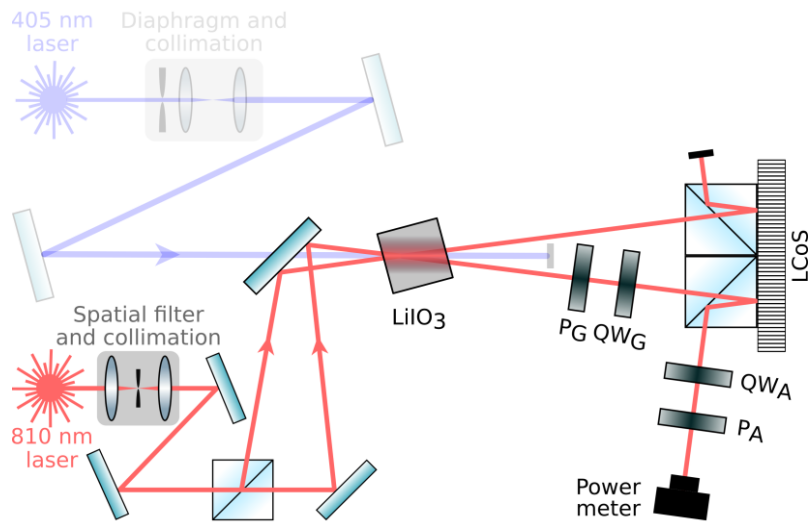

Fig. 7. Schematic of the experimental setup used to measure the Mueller matrices of a LCoS display. A 810 nm laser beam mimics the down-converted photons from a $LiIO_3$ nonlinear crystal. Light is analyzed by a Mueller polarimeter (polarizers $P_G$ and $P_A$, and quarter-wave plates $QW_G$ and $QW_A$) and detected by a power meter.

As it can be seen in that figure, the curves are very similar to each other. The slight differences between them can be explained by misalignments in the optical setup, mainly due to the fact that the laser beams do not perfectly follow the same path as the SPDC photons. Also, the SPDC measurements took several hours to be performed, so changes in the detection rate may be explained by fluctuations in the intensity of the pump, and changes in temperature and humidity that may modify the efficiency of the nonlinear crystal to down-convert photons. This agreement is confirmed through the Pearson correlation coefficient between the curves of every matrix element, which can be found in Fig. 6 (b). The first row and column are not well correlated, but it can be attributed to the fact that the value of those elements is close to zero for every gray level, as seen in Fig. 6 (a). The remaining elements are the ones that change with the gray level, and their mean correlation coefficients are 0.992 with a standard deviation of 0.010.

## 4. Conclusions

In this work, we addressed two assumptions usually made when using SLMs for the control of quantum states of light. In particular, we studied the ability of an LCoS display to preserve spatial momentum-position correlations of twin photons, and the possible differences in its Mueller matrices when changing the light intensity regime. In both cases, the source of single photons was a nonlinear crystal pumped by a laser diode.

In order to study the correlations, two experimental setups were used: one in which photons could propagate freely, and one were they reflected onto the LCoS display. By doing this, we were able to analyze the position and momentum distributions of the down-converted photons. By means of their variances product the position-momentum correlation was estimated, confirming that it doesn't change within the experimental errors due to the presence of the display, nor with the set gray level. This is important, since it means that this kind of spatial light modulators can be used to control quantum states without deteriorating this key property of the photon pairs.

To obtain the Mueller matrices, a single setup was used, but illuminated with two different sources: a $LiIO_3$ nonlinear crystal, and a laser of the same wavelength as the down-converted photons. By measuring the intensity of the beam through the different configurations of a Mueller polarimeter, the Mueller matrices of the gray levels of the display were estimated. The results confirmed that those matrices are the same after considering small deviations in the optical setup. This is also important, as a characterization of the SLM is necessary, and measuring the Mueller matrix with single photons is a time-consuming task that requires precise alignment, which can be saved by estimating them using intense light.

These two analyses confirm that LCDs are a valuable choice for several experimental applications in optics, and particularly for the use of single photons in quantum optics experiments.

**Acknowledgements**

This work was supported by CONICET (PIP 11220080103047) and SCyT-UBA (UBACyT 20020170100564BA). Sebastián Bordakevich and Dudbil Pabón were supported by CONICET fellowships.